# A new method for imaging nuclear threats using cosmic ray muons

C. L. Morris, Jeffrey Bacon, Konstantin Borozdin, Haruo Miyadera, John Perry, Evan Rose, Scott Watson, and Tim White

*Los Alamos National Laboratory, Los Alamos, NM, 87545 USA.*

Derek Aberle, J. Andrew Green and George G. McDuff

*National Security Technologies, Los Alamos, NM, 87544 USA.*

Zarija Lukić

*Lawrence Berkeley National Laboratory, Berkeley, CA, 94720 USA.*

Edward C. Milner

*Southern Methodist University, Dallas, TX, 75205, USA.*

**Abstract** Muon tomography is a technique that uses cosmic ray muons to generate three dimensional images of volumes using information contained in the Coulomb scattering of the muons. Advantages of this technique are the ability of cosmic rays to penetrate significant overburden and the absence of any additional dose delivered to subjects under study above the natural cosmic ray flux. Disadvantages include the relatively long exposure times and poor position resolution and complex algorithms needed for reconstruction. Here we demonstrate a new method for obtaining improved position resolution and statistical precision for objects with spherical symmetry.

## Introduction

Cosmic ray muon scattering measurements provide a method for imaging the internal structure of objects using the information contained in the naturally occurring cosmic ray flux found at the surface of the earth[1-3]. The use of the signal from Coulomb scattering enables three dimensional imaging of complex scenes with high sensitivity to objects composed of materials with high atomic charge (Z). This makes this technique uniquely suitable for detecting and measuring the properties of atomic explosives that may be inaccessible with other techniques.

If a nuclear material has been detected it is important to be able to measure details of its construction in order to correctly evaluate the threat. The flux of cosmic rays is low so it takes very long exposures to produce images with high resolution. In this paper we show how one can take advantage of spherical symmetry to improve the statistical precision of muon imaging.

## Three Dimensional Imaging

Although cosmic rays are highly penetrating and can image through considerable overburden, the flux is limited. The times required detecting the presence of quantities of uranium or plutonium necessary to create a nuclear explosion are on the order of minutes, and the times needed to image these devices with ~2 cm resolution are on the order of hours.

Imaging with cosmic rays is based on measuring the multiple Coulomb scattering of the muons. The dominant part of the multiple scattering polar-angulardistribution is Gaussian:

$$\frac{dN}{d\theta} = \frac{N}{2\pi\theta_0^2} e^{-\frac{\theta^2}{2\theta_0^2}}, \qquad 1)$$

the Fermi approximation, where $\theta$ is the polar angle and $\theta_0$ is the mean multiple scattering angle, which is given approximately by:

$$\theta_0 = \frac{14.1\,\text{MeV}}{p\beta}\sqrt{\frac{l}{X_0}} \qquad 2)$$

The muon momentum and velocity are $p$ and $\beta$, respectively, $l$ is the material thickness, and $X_0$ is the radiation length for the material. This equation needs to be convolved with the cosmic ray momentum spectrum in order to describe the angular distribution.

There are several algorithms for generating tomographic images using the input and the output trajectories of the cosmic rays described by Schultz et al.[2] Here we use a very simple method with a scene composed of voxels . For each voxel a one-dimensional histogram of scattering angle in 1 mrad steps from 0 to 200 mrad is created. The histogram is incremented for every cosmic ray for which the incident cosmic ray intercepts the voxel and for which the distance between the input and the output cosmic rays in the plane of the voxel is less than 2 cm. For large scattering angles this requirement associates measured scattering events with defined voxels in which the most of the scattering occurred.

The scattering distribution for each voxel is fitted with a model that uses seven momentum groups,[4] $p_i$, to approximate the muon spectrum with,

$$\frac{dN}{d\theta} = \sin(\theta) \sum \frac{A_i}{\theta_{0i}^2} e^{-\frac{\theta^2}{2\theta_{0i}^2}}$$

$$\theta_{0i} = \frac{14.1}{p_i} \sqrt{\frac{l}{X_0}}$$

3)

to approximate the muon energy distribution.

The model has been calibrated with data taken through three thicknesses of lead, 5.08, 10.16 and 15.24 cm. The amplitudes, $A_i$, of each energy group, as well as the intrinsic angular resolution and a fixed number of radiation lengths due to the drift tubes and other structural elements of the muon detectors were fitted to minimize the logarithm of the likelihood function assuming the data were describe by a Poison distribution. This model does not account for changes in the shape of the muon spectrum due to stopping. A maximum likelihood fit to the set of lead data is shown in **Error! Reference source not found.**. Also shown is the decomposition of one of the data sets into its momentum groups.

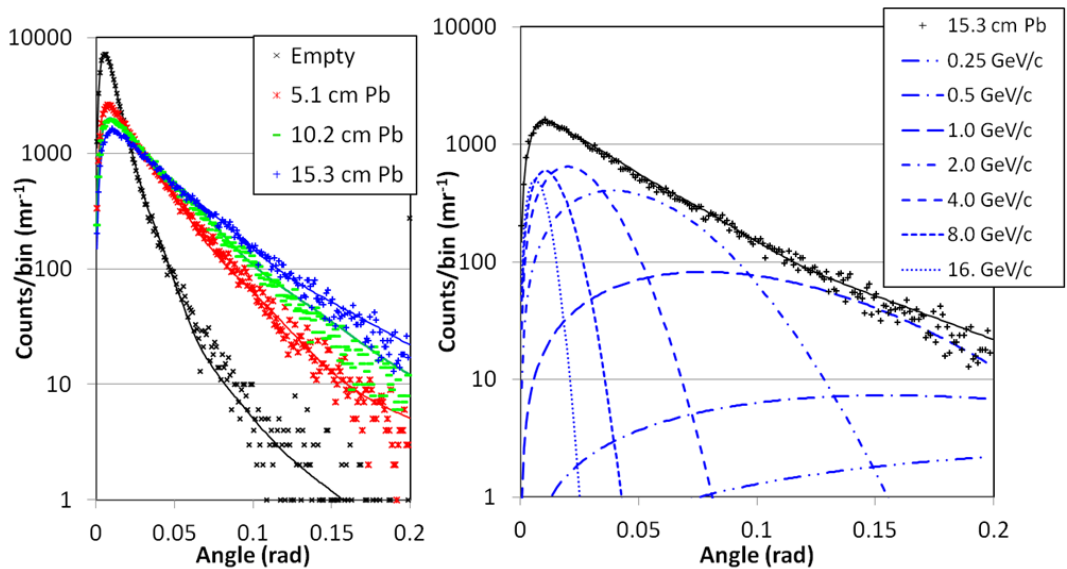

Figure 1) Multi-group fit to the lead calibration data. On the left is a plot of the fits to different thicknesses of lead, the plot on the right shows the decomposition of the fit into its multiple Gaussian components.

Images were constructed by fitting the angular distribution for each voxel to obtain the average number of radiation lengths of material that the ensemble of histogram entries has traversed. This reconstruction algorithm is scalable to large data sets, is simple to compute, and provides near optimal use of the scattering information. However, it doesn't optimally use the vertex information.

We have imaged several spherically symmetric objects: nested spherical shells of copper and tantalum, the copper shell alone, and a hollow lead ball. The outer radii were 6.5, 4.5, and 10 cm and the inner radii were 4.5, 1, and 2.5 cm for the copper, tantalum and lead shells respectively. Cartesian slices through the three dimensional tomograms, centered on the object, are shown in Figure 2.

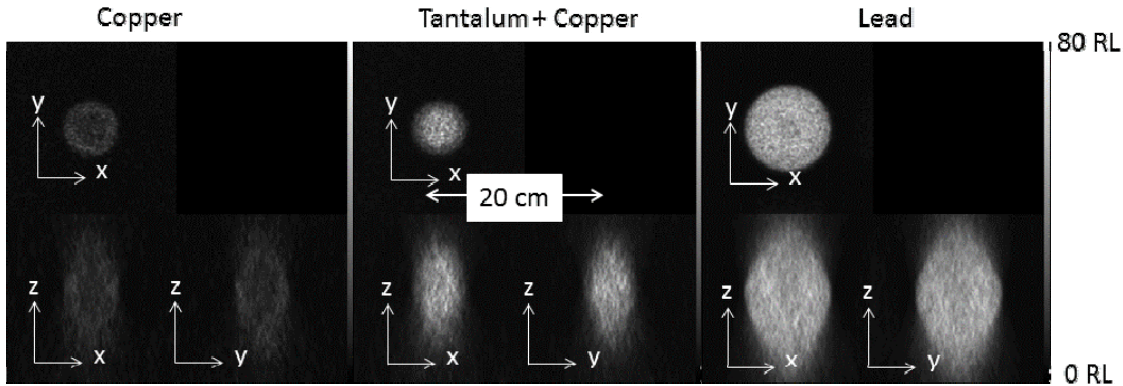

Figure 2) Cartesian slices through the tomographs of the three objects. They are presented on the same position and grey scales. The grey scale is linear between 0 and 80 radiation lengths from black to white respectively.

## One Dimension Imaging

The center of each object ($x_c, y_c, z_c$) was estimated by finding the centroid of the signal from the object using the Cartesian slices shown in Figure 2. The were used for a one-dimensional reconstruction. A trajectory was defined by the line **x**(s) where:

$$x(s) = x_0 + x's$$
$$y(s) = y_0 + y's \qquad \qquad 4)$$
$$z(s) = z_0 + z's$$

Here ($x_0$, $y_0$, $x_0$) is a point on the line with direction cosines (x', y', x'). The point of minimum distance between ($x_c$, $y_c$, $z_c$) and ($x_0$, $y_0$, $x_0$) is given by s=$s_0$:

$$s_0 = \frac{(x_c - x_0)x' + (y_c - y_0)y' + (z_c - z_0)z'}{x'^2 + y'^2 + z'^2}. \qquad 5)$$

The radius of closest approach is:

$$r_0 = \sqrt{(x(s_0) - x_c)^2 + (y(s_0) - y_c)^2 + (z(s_0) - z_c)^2}. \qquad 6)$$

A two dimensional histogram of scattering angle versus radius is shown in Figure 3.

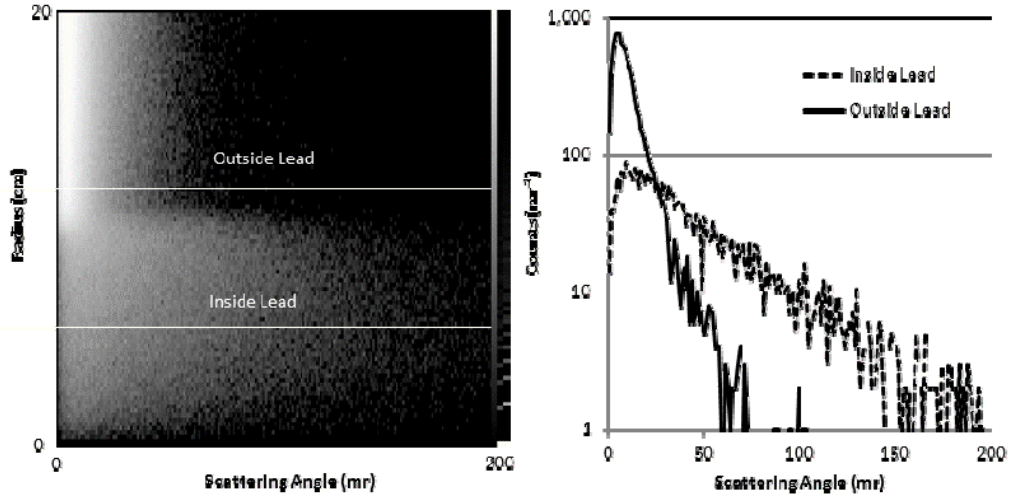

**Figure 3)** Left) scattering angle vs. radius for the lead spherical shell. The grey scale is proportional to the logarithm of the number of counts per bin. On the right are plots of counts vs. scattering angle taken along the lines shown in the plot on the left.

A spherically symmetric object can be described by a set of shells at $r_i$ with thickness $dr = (r_{i+1} - r_{i-1})/2$ and of a material with radiation length, $X_{0i}$, and weighted density $\rho_{vi}/X_{0i}$. The radiation length weighted path length, $L_i$, as a function of $r_i$ can be obtained from the data shown in the 2-dimensional histogram by using the multi-group fitting technique described above (Equation 3) for each radial bin. The fit to the lead data shown in Figure 1 has been corrected by 12% to account for the average $1/\cos(\theta)$ increase in thickness of the lead in the planar geometry and then used here.

The $L_i$ are related to the set of radiation length weighted volume densities, $\rho_{vj}$, by a path length vector $P_{ij}$ (see Figure 4):

$$L_i = P_{ij} \frac{\rho_{vj}}{X_{0j}}.  \qquad 7)$$

The path length vector is the length a particle at $r_i$ traverses through shell j:

$$P_{i,j} = 0 \quad \text{for } i > j$$

$$P_{i,j} = 2\sqrt{\left(r_j + \frac{dr}{2}\right)^2 - r_j^2} \quad \text{for } i = j \qquad . \qquad 8)$$

$$P_{i,j} = 2\sqrt{\left(r_j + \frac{dr}{2}\right)^2 - r_j^2} - 2\sqrt{\left(r_{j-1} + \frac{dr}{2}\right)^2 - r_{j-1}^2} \quad \text{for } i < j$$

This can be solved for the $\rho_{vj}/X_{0j}$ using the regularization techniques described in Press.[5] This technique dampens the on-axis noise that arises in conventional Abel inversions[6] which is important here because of the poor statistics, especially at small r.

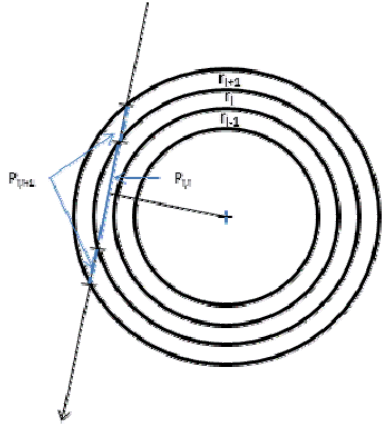

**Figure 4) An Illustration of the path-length matrix, $P_{i,j}$.**

The results for the three objects, each with 24 hours of exposure, are shown in Figure 5. The $\rho_{vj}/X_{0i}$ for each of the materials studied here, copper, tantalum, and lead, are with 10% of the tabulated values.[7] One can easily distinguish the void inside each of the shells, even the 1 cm radius void at the center of the tantalum shell. An analysis of the width of the edges in Figure 5 give a position resolution of 3 mm.

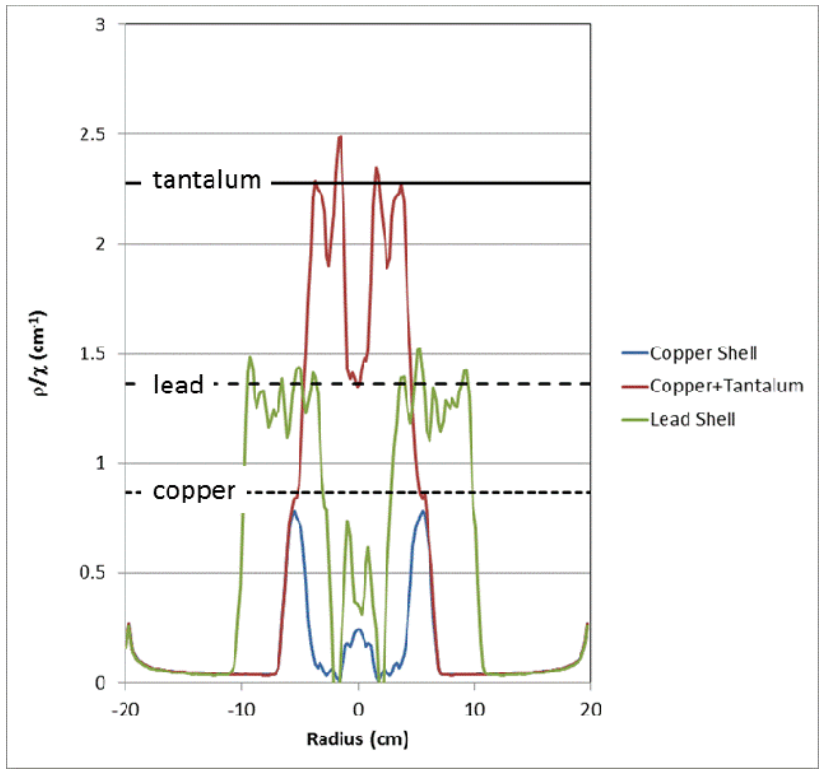

**Figure 5) Radiation length weighted density vs radius. The radii have been mirrored around r=0 (the data at negative *r* are the same as the data for positive *r*). Horizontal lines show the tabulated value of ρ/X.[7]**

It is worth noting that these objects are difficult to study with conventional x-ray radiography. While the cavity can be observed with relatively crude collimation techniques, quantifying the cavity density takes sophisticated anti-scatter techniques.[8]

## Conclusion

Cosmic ray scattering data were taken on a set of spherically symmetric objects. The data were analyzed assuming spherical symmetry. The data were stored in a two dimensional histogram of scattering angle vs. radius. The angular distributions were fitted by a sum of Gaussians whose amplitudes were fixed by fits to data taken on a set of planar objects. This resulted in one-dimensional plots of thicknesses in radiation lengths for each of the objects. These were analyzed with a regularized Abel inversion technique yielding radiation-length-weighted volume densities. These results allowed a quantitative evaluation of the material composition of the objects.

## Acknowledgements

We would like to acknowledge help from Dave Schwellenbach and Wendi Dreesen setting up the hardware and software that has enabled these measurements. This work was supported in part by the United States Department of Energy, the United States Department of State, and the Defense Threat Reduction Agency of the United States Department of Defense.

LA-UR-13-22644